# Quantum tunneling time

P.C.W. Davies

*Australian Centre for Astrobiology, Macquarie University, New South Wales
Australia 2109*

**ABSTRACT**

A simple model of a quantum clock is applied to the old and controversial problem of how long a particle takes to tunnel through a quantum barrier. The model I employ has the advantage of yielding sensible results for energy eigenstates, and does not require the use of time-dependant wave packets. Although the treatment does not forbid superluminal tunneling velocities, there is no implication of faster-than-light signaling because only the transit duration is measurable, not the absolute time of transit. A comparison is given with the weak-measurement post-selection calculations of Steinberg.

## 1. INTRODUCTION

Recent experiments on quantum tunneling have re-ignited the longstanding debate over how long a particle takes to tunnel through a barrier [1]. Naïve calculations suggest that faster-than-light tunneling is possible, while some experiments have superficially suggested that such a phenomenon might have been observed [2]. However, it has been argued that *information* does not exceed the speed of light in these experiments, so that relativistic causality remains intact [3].

The analysis of tunneling time is complicated because time plays an unusual and subtle role in quantum mechanics. Unlike position, time is not usually treated as an operator; rather it is a parameter. Consequently, the energy-time uncertainty principle does not enjoy the unassailable central position in the theory as does the position-momentum uncertainty principle. This leads to considerable ambiguity when it comes to the measurement of the duration between quantum events.

Attempts to define tunnelling time have led to an extensive and confused literature [4]. Most theoretical treatments focus on the behaviour of a wave packet as it traverses a square barrier. However, such a barrier is dispersive, so the packet is disrupted by the experience. Also, interference between parts of the wave packet reflected from the barrier and parts still approaching further complicates matters.

A simple heuristic argument to estimate the tunnelling time goes as follows. To surmount a square barrier of height $V$, a particle with energy $E$ must 'borrow' an amount of energy $V - E$. According to the uncertainty principle, this must be 'repaid' after a time $T = 1/(V - E)$ in units with $\hbar = 1$. This provides a crude upper bound for the tunneling time. If the width of the barrier is *a*, then the effective speed of the particle during the tunneling process must exceed $a(V - E)$. As *a* can be made as large as we please, there is no upper bound on this effective velocity. In particular, it may exceed the speed of light, in apparent violation of relativistic causality. Moreover, the above expression has the odd feature that as the height *V* of the potential hill is increased, so the tunneling time decreases, i.e. the more repulsive the potential, the *faster* the particle moves in the forward direction!

A natural way to approach the problem is to introduce some sort of clock that is coupled to the particle, and to define the tunneling time in terms of the change in the clock variable from the time that the particle reaches the barrier to the time it emerges. It is then possible to define the expectation time for the tunneling event in terms of the expectation value of the clock variable, which *is* a quantum observable. Several suggestions for quantum clocks have appeared in the literature [5]. In this paper I restrict attention to an early proposal for a quantum clock by Salecker and Wigner [6], and later elaborated by Peres [7]. This model deals not with moving wave packets or other time-dependent states, but with stationary states for which the time enters as a changing phase: $e^{iEt}$. The quantum clock essentially measures the change in phase, resulting in a duration known as the *phase time*. Because one is dealing with stationary states, the clock measures only time *differences* between two events, not the absolute time of either event.

This is a key point. Suppose one tries to measure the tunneling time by first measuring the time at which the particle arrives at the leading edge of the potential hill, then measuring the time at which it emerges on the remote side, and taking the difference. The act of observing the particle at the first position collapses the wave function to a position eigenstate and introduces arbitrary uncertainty into the momentum, so that the second measurement is upset. If, on the other hand, one forsakes knowledge of the absolute time of passage of the particle, but requires only the duration for the particle to pass between two fixed points in space, then only a single measurement is required and there is no large unpredictable disturbance. To achieve this, the particle is coupled (weakly) to a quantum clock. The coupling is chosen to be non-zero only when the particle's position lies within a given spatial interval (e.g. the potential hill). Initially the clock pointer is set to zero. After a long time, when the particle has traversed the spatial region of interest with high probability, the position of the clock pointer is measured. The change in position yields the expectation value for the time of flight of the particle between the two fixed points. Full details are given in [7] and will not be repeated here.

To illustrate the method, consider a free particle of mass *m* and energy *E* moving to the right in one space dimension, in a momentum eigenstate $e^{ikx}$ described by the time-independent Schrödinger equation. The expectation value for the time of flight between two points separated by a distance $\Delta$ is found as follows. First compute the phase difference $\delta(E)$ in the wave function between the two points; this is $\delta(E) = k\Delta$ where $k \equiv (2mE)^{1/2}$. Next replace *E* by $E + \varepsilon$ where $\varepsilon$ is the coupling energy between the particle and the clock, treated in first order perturbation theory. Now expand $\delta(E + \varepsilon)$ to first order in $\varepsilon$. The coefficient, $\delta'(E)$, is *T*, the required expectation value of the time for the particle to traverse the distance $\Delta$. In the above case $T = \delta'(E) = m\Delta/(2mE)^{1/2}$. Defining the classical velocity $v = (2mE)^{1/2}$, the expected time of flight *T* is seen to be identical to the classical result $\Delta/v$.

## II. POTENTIAL STEP

As a less elementary example, consider the particle scattering from a potential step of height $V$ situated at $x = 0$. The stationary state wave function has space-dependent part

$$e^{ikx} + Ae^{-ikx} \qquad x < 0,$$

$$Be^{-px} \qquad x > 0. \qquad (1)$$

where $k \equiv (2mE)^{1/2}$ and $p = [2m(V - E)]^{1/2}$. It turns out that the overall normalization factor does not affect the result, so it will be omitted. Suppose we require the expectation value for the time of flight of the particle to travel from $x = -b$ to the barrier and back again. At $x = -b$ the phase of the incident portion of the wave function is $-kb$. Now compute the phase of the reflected portion at $x = -b$. Using continuity of the wave function and its first derivative at the step, we solve for $A$ and find for the said phase the value $kb + \alpha$ where

$$\alpha = \arctan(\text{Im } A/\text{Re } A) \qquad (2)$$

and

$$A = -(p + ik)/(p - ik). \qquad (3)$$

Thus $\delta(E) = 2kb + \alpha$ and we find by differentiating with respect to $E$ that

$$T = 2b/v + 2m/kp = 2(b + d)/v \qquad (4)$$

where $d \equiv 1/p$ is the expectation value for the penetration depth of the particle into the potential step.

The result Eq. (4) has an intuitively simple interpretation. The term $2b/v$ is the time of flight from $x = -b$ to the potential step at $x = 0$ and back again, at the classical velocity $v$. The term $2d/v$ represents (the expectation value of) the additional duration of sojourn of the particle in the classically forbidden region beneath the potential step, and can be interpreted as if the particle moves with the classical velocity $v$ for a distance $d$ equal to the average penetration depth, and back again. Thus the effective distance from $x = -b$ to the step is increased from the classical distance $b$ to $b + d$. Note that if $V \to \infty$ then $d$ vanishes, so an infinite potential step yields instantaneous reflection. On the other hand, as $E \to V$, $p \to 0$ and the sojourn time beneath the step diverges: the particle takes an infinite time to bounce back.

In the case that $E > V$, $p$ is imaginary, $A$ is real and $\alpha = 0$. The round-trip time therefore reduces to the classical result $2b/v$. The reflection from the step is instantaneous in this case too, even when $E \to V$ from above. There would thus appear to be an infinite discontinuity in the reflection time at $E = V$. However, we must be cautious. The method of computation demands that we expand functions of $E - \varepsilon$ and $V - E - \varepsilon$ in powers of $\varepsilon$ and treat $\varepsilon$ as small. This procedure is clearly untrustworthy near both $E = 0$ and $E = V$. I shall return to this problem later.

One may also compute the time for the particle to go from $x = -b$ to, say, $x = b' > 0$ by examining the phase of the wave function in the region $x > 0$ (the $B$-dependent term in Eq. (1)). If $E < V$, the method suggests an imaginary phase shift $-pb'$, implying an imaginary time. If $E > V$, we obtain $b/v + b'/v'$, where $v' = [2(E - V)/m]^{1/2}$ is the classical velocity above the step. Evidently we can directly patch together the time of flight before the step with that at the reduced velocity after the step.

### III. POTENTIAL HILL

I now treat the case of principal interest: a particle that tunnels through a square potential hill given by $V = $ constant $> 0$ in the interval $[0, a]$ and zero elsewhere. The wave function is

$$e^{ikx} + Ae^{-ikx} \qquad x < 0$$

$$Be^{px} + Ce^{-px} \qquad 0 < x < a$$

$$De^{ikx} \qquad x > a. \qquad (5)$$

The phase of the incident part of the wave function at $x = 0$ ('entering the tunnel') is 0. The phase of the emergent wave function at $x = a$ ('leaving the tunnel') is given by the phase of $De^{ikx}$. Using continuity of the wave function and its derivative at $x = 0$ and $x = a$, the phase change is found to be

$$\delta(E) = \arctan\{[(p^2 - k^2)/2kp]\tanh(pa)\}. \qquad (6)$$

Differentiation then yields for the expectation value of the tunneling time

$$T = 2m\{k(p^2 - k^2)a + [(p^2 + k^2)^2/2kp]\sinh 2pa\}/[(p^2 + k^2)^2\cosh^2 pa - (p^2 - k^2)^2]. \qquad (7)$$

As a check, we note that when $V = 0$, $p = ik$ and $T = ma/k = a/v$ as expected.

In the special case $E = V/2$, the right hand side of Eq. (7) reduces to $\tanh(ka)/E$. For small $a$, it reduces to

$$(ma/2k)(3 + p^2/k^2) \qquad (8)$$

which $\to 0$ as $a \to 0$, as expected. If we define the *effective velocity* of the particle to be $v_{\text{eff}} \equiv a/T$, then for small $a$

$$v_{\text{eff}} \approx 2v/(2 + V/E), \qquad (9)$$

where $v = k/m = (2mE)^{1/2}$ is the classical velocity of the particle outside the potential hill. Note that $v_{\text{eff}} < v$ in this limit: thin potential hills slow the particle down, as one might be led to believe on classical grounds. For $E = V/2$, $v_{\text{eff}} = v/2$. In the case that $V \gg E$, the limit used above breaks down. A case of interest is a delta-function potential hill, where $Va^2 = $ constant as $V \to \infty$ and $a \to 0$. Returning to Eq. (7) and applying these limits, one finds that $T \to 0$. There is no problem here about reflected waves slowing the particle as it approaches the barrier. This result confirms the work of Aharonov, Erez and Reznik [8], who find $T = 0$ for the tunnelling time through an array of delta function potential hills.

By contrast to the slowing effect found above, thick hills serve to speed the particle up, i.e $v_{eff} > v$ in this case. Taking the limit $a \to \infty$ in Eq. (7) we see that the tunneling time approaches the constant value

$$2m/kp = [E(V - E)]^{-\frac{1}{2}}. \tag{10}$$

This is similar to the result found from the naive argument mentioned in section I. The right hand side of Eq. (10) is reminiscent of the energy that can be 'borrowed' for a time $T$ according to Heisenberg's uncertainty principle, but with the interesting difference that the 'borrowing' requirement is not simply $V - E$, but the harmonic mean of this quantity and $E$. The tunneling time is minimized for $E = V/2$ and in this case we do have $T = 1/E$. Note that Eq. (10) is also equal to the second term on the right hand side of Eq. (4), the sojourn time inside a potential step. We shall see below that this is a special limit of the general result that the expectation time for a particle to reflect back from the potential barrier is the *same* as the expectation time for it to penetrate the barrier.

The effective velocity under the barrier is

$$v_{eff} = aE^{\frac{1}{2}}(V - E)^{\frac{1}{2}} \tag{11}$$

which rises without limit as $a \to \infty$. In particular, $v_{eff}$ exceeds the speed of light $c$ when

$$pa > 2mc/k = 2(\text{de Broglie wavelength})/(\text{Compton wavelength}). \tag{12}$$

However, for thick barriers the transmission probability is very small. To estimate it, first note that if the approaching particle is to remain non-relativistic (as assumed in the treatment given here), then the right hand side of Eq. (12) must be $\gg 1$, which implies $pa \gg 1$. In this limit the transmission probability approximates to

$$16(E/V^2)(V - E)e^{-2pa}. \tag{13}$$

Consider the example of $E = V/2 = mc^2/8$. Taking $pa = 2mc/k$ (corresponding to the onset of superluminal propagation), the barrier penetration probability is then $4e^{-8} \approx 10^{-3}$. Although small, this number is by no means negligible, and we have to confront the consequences for causality if it is indeed the case that the occasional particle can tunnel faster than light.

A violation of causality will come about if observer $A$ can send information to an observer $B$ a distance $d$ away such that it arrives before a time $d/c$ has elapsed. Could $A$ use an electron to encode this information, and arrange for it to tunnel through a barrier to $B$ in the knowledge that, albeit only occasionally, $B$ will get to receive the electron before a time $d/c$? I believe the answer to be no. To achieve physical causality violation, $A$ must be able to determine the moment of transmission of the information. But as we have seen, the model system discussed here can determine only the time *difference* between the moment of 'transmission' and 'reception' of the particle - not the absolute time of transmission. If $d/c - T = \Delta t$, say, then to qualify as causally relevant, the signalling process must be controlled to a fidelity $< \Delta t$; but this is not possible in the present model. Any faster-than-light propagation would therefore be fortuitous - entirely random and uncontrollable. Tunnelling may violate the spirit of relativity, but it does not seem to violate the letter.

## IV. MEASUREMENT UNCERTAINTY

The model clock used here is a quantum system, and is therefore subject to quantum uncertainty in its operation, which in turn implies an uncertainty in the deduced tunneling time. As shown by Peres [7], back action of the clock's dynamics on the particle's motion, which persists throughout the 'experiment,' will limit the resolution of this model clock. In particular, it is unreliable when $E \to 0$ or $E \to V$.

The resolution of the clock is limited by the assumption that $|E|$ and $|V - E|$ are $\gg \varepsilon$. The energy-time uncertainty relation applied to the clock variables then suggests that the clock pointer will have an uncertainty corresponding to a time $\tau \approx 1/\varepsilon \gg 1/E$. But the tunneling time as illustrated by, say, the asymptotic value Eq. (10), is itself of order $1/E$. Hence the quantum uncertainty in measuring the tunneling time $T$ is of the same order as the expectation value of the tunneling time. This is no surprise, as any limitation in measurement resolution will have this general form on dimensional grounds.

The disturbance on the particle's motion caused by the back action can be reduced by making the coupling weaker, but at the expense of introducing greater uncertainty in the measurement of the clock reading. An alternative strategy to reduce the uncertainty is to use a clock that is not continuously coupled to the particle. This could be achieved by placing the clock in a metastable state, and then using the arrival of the particle at the leading edge of the barrier to merely trigger the operation of the clock via a momentary interaction. This sort of device has been studied by Oppenheim, Reznik and Unruh[9]. Perhaps surprisingly, it does not result in a reduction in the overall uncertainty. The reason for this is that the sharply-localized potential associated with the triggering device reflects some of the wave function, and attempts to mitigate this back-action effect (for example, by boosting the energy of the particle just before the barrier) serve only to introduce additional uncertainties.

Thus there seems to be an irreducible uncertainty in the measurement of the tunneling time that is comparable to the tunneling time itself. At first sight this appears to cast doubt over the usefulness of the foregoing results. However, as shown by Aharonov et. el. [10], by performing measurements on a large ensemble of identical systems, the spread in results can be drastically narrowed, even in cases where the uncertainty for a single measurement exceeds the expected value. This is the theory of weak measurement. Applied to the problem of the Peres clock and tunneling time, weak measurement theory implies that, interpreted in an ensemble sense, the results of the foregoing sections are physically meaningful, in spite of the intrinsic uncertainty in the operation of the clock.

Weak measurement theory is often combined with post-selection, whereby a final sub-ensemble is extracted corresponding to the state of interest. In the case of tunneling, this sub-ensemble will include only those particles that penetrate the barrier and move to the right. Steinberg [11] has computed the tunneling time using this approach, by evaluating an expectation value for a projection operator corresponding to the time the particle is inside the barrier, in the limit that the measuring device interacts only exceedingly weakly with the particle. The resulting expression is complex. Its real part corresponds to the expectation value of the tunneling time, the imaginary part to the back action of the measuring device on the particle. These respective parts are related in a rather transparent manner to other proposed definitions of the tunneling time. For example, the real part is identical to the so-called dwell time, which is defined as the probability of finding the particle inside the barrier divided by the incoming flux.

Steinberg's result for the tunneling time expectation value is

$$T_s = 2m\{k(p^2 - k^2)a + k(p^2 + k^2)/2p]\sinh 2pa\}/[(p^2 + k^2)^2\cosh^2 pa - (p^2 - k^2)^2]. \quad (14)$$

which should be compared to Eq. (7). The two durations are very similar, but not identical. One finds

$$T/T_s = V/E + (1 - V/E)/[1 + (V\sinh 2pa)/2(V - 2E)pa]. \quad (15)$$

In the free-particle limit $V \to 0$, $T \to T_s$. In general $T_s < T$ for $E < V$. In the limit of large barrier width $a$

$$T_s \to (E/V)T = (E/V)[E(E - V)]^{-1/2}. \quad (16)$$

Note that both $T$ and $T_s$ diverge as $E \to V$, but in the limit $E \to 0$ the behaviour is very different:

$$T \to \infty \quad (17)$$

$$T_s \to 0, \quad (18)$$

the latter result implying the curious property that, as the approach velocity of the particle falls, so the tunneling velocity rises; in the limit $v \to 0$ the post-selected tunneling velocity *diverges*.

## V. OTHER RESULTS

One may use the Peres clock model to calculate some other transit times of interest. Consider, for example, the time between incidence and reflection from the leading face of the hill at $x = 0$. This may be computed by examining the phase of $A$ in Eq. (5). One finds for the phase change

$$\delta(E) = \text{arccot}\{[p^2 - k^2]/2kp]\tanh(pa)\}. \quad (19)$$

But the derivative of Eq. (19) is identical to that of Eq. (6), so the sojourn time inside the hill is the same, whether the particles are transmitted or reflected.

Thus the tunneling time for both transmission and reflection are the same. It has been argued [4] that tunneling times for transmission and reflection should satisfy the relation

$$T_D = P_t T_t + P_r T_r \qquad (20)$$

where $T_D$ is the dwell time, $P_t$ ($P_r$) the probability of transmission (reflection) and $T_t$ ($T_r$) the corresponding tunneling times. Equation (20) is satisfied, for example, by Steinberg's definition, but not by the one used in this paper. However, Landauer and Martin [1] have argued strongly against Eq. (20) as an inappropriate criterion.

The analysis given in this paper to investigate tunnelling may also be used to derive results for one-dimensional scattering, by putting $E > V$. Defining $\beta \equiv [2m(E - V)]^{1/2}$, Eq. (7) becomes

$$T = 2m\{k(k^2 + \beta^2)a + (k^2 - \beta^2)^2/2k\beta]\sin 2\beta a\}/[(k^2 - \beta^2)^2 \cos^2 \beta a - (k^2 + \beta^2)^2]. \qquad (21)$$

For small $a$ the right hand side of Eq. (21) reduces to

$$(ma/2k)(3 - \beta^2/k^2) \qquad (22)$$

with a corresponding effective velocity given by Eq. (9). Again, $v_{\text{eff}} < v$, although $v_{\text{eff}} > [2m(E - V)]^{1/2}$, the classical velocity over the barrier. So although the repulsive potential slows the particle, it does not do so as much as in the classical case.

Now consider the opposite limit of large $a$. The denominator on the right hand side of Eq. (21) can never vanish, while the sine function in the numerator is bounded by $[-1,1]$. Thus

$$T \approx 2mka/(k^2 + \beta^2) = ka/(2E - V). \qquad (23)$$

By contrast to the result for the tunneling case, the right hand side of Eq. (23) is proportional to $a$ even for large $a$. The effective velocity $(2E - V)/k$ therefore always remains less than $c$ when the particle passes *over* the barrier: only tunneling events lead to superluminal velocities. For large $E$, $v_{\text{eff}} \to v$, but for particles that just clear the barrier, $E \approx V$, $v_{\text{eff}} \approx v/2$.

Special interest attaches to the case of resonance transmission, when $\beta a = n\pi$, and $P_t = 1$. Then Eq. (21) simplifies:

$$T = ma(k^2 + \beta^2)/2k\beta^2 \qquad (24)$$

$$v_{\text{eff}} = v(E - V)/(E - V/2). \qquad (25)$$

In this case $v_{\text{eff}}$ does approach 0 as $E \to V$, as it would in the classical case. The effective velocity, however, has a very different energy dependence from the classical expression. For the case of anti-resonance, where $\cos(\beta a) = 0$, Eq. (23) becomes exact, and

$$v_{\text{eff}} = \tfrac{1}{2}(v + v') \qquad (26)$$

- the average of the classical velocities outside and over the barrier.

Note that because the reflection and transmission expectation times are equal, there is always a reflection delay, or sojourn in the region $x > 0$, even in the case that $E > V$. This is in contrast to the single potential step, where reflection is instantaneous if $E > V$. The difference has a natural interpretation. In the case of the potential hill, reflection may take place from both the leading and remote faces of the hill. The actual reflections may be instantaneous, but in the case that the particle reflects from the far edge $x = a$ there will be a delay due to the travel time across the top of the hill. The expectation value will therefore include this delay, and the result is consistent with one half the flux being reflected from each edge.

Finally, it is worth noting that the above analysis applies to the case of scattering from a potential well, $V < 0$. For resonant scattering, where a bound state exists just below the top of the well, the scattering cross-section rises sharply at low energy. This is accompanied by a rise in the scattering time $T$. However, although the effect is explicit in a calculation of the dwell time, $T_D$, it is masked in the case of the Peres clock, because $T$ diverges anyway as $E \to 0$.

## VI. CONCLUSION

I have shown by use of a simple model that sensible and consistent expressions may be derived for the expectation value of the time for a non-relativistic particle in an energy and momentum eigenstate to pass between two points, so long as the absolute time of passage is not required. The points may be separated by regions that include a variety of potentials, including a square potential barrier. In the latter case the tunneling time is given by a credible expression, which approaches a constant for thick barriers, implying an issue concerning superluminal propagation. However, I have argued that physical causality is not violated.

In this paper I have restricted the discussion to simple square barriers. It is of interest to consider tunneling into other types of potential too. An important example is the uniform gravitational potential $V(x) = mgx$. I have discussed this problem in detail in another publication [12].

## ACKNOWLEDGMENTS

I am grateful to Aephraim Steinberg for helpful comments.